\documentclass[twocolumn]{aastex63}
\usepackage{amsmath}
\shorttitle{Kinetic model of large-amplitude oscillations}
\shortauthors{Cruz et al.}
\graphicspath{{./}{figures/}}

\begin{document}

\title{Kinetic model of large-amplitude oscillations in neutron star pair cascades}

\correspondingauthor{F\'{a}bio Cruz}
\email{fabio.cruz@tecnico.ulisboa.pt}

\author[0000-0003-0761-6628]{F\'{a}bio Cruz}
\affiliation{GoLP/Instituto de Plasmas e Fus\~{a}o Nuclear, \\ Instituto Superior T\'{e}cnico, Universidade de Lisboa, 1049-001 Lisboa, Portugal}

\author[0000-0002-0045-389X]{Thomas Grismayer}
\affiliation{GoLP/Instituto de Plasmas e Fus\~{a}o Nuclear, \\ Instituto Superior T\'{e}cnico, Universidade de Lisboa, 1049-001 Lisboa, Portugal}

\author[0000-0003-2906-924X]{Luis O. Silva}
\affiliation{GoLP/Instituto de Plasmas e Fus\~{a}o Nuclear, \\ Instituto Superior T\'{e}cnico, Universidade de Lisboa, 1049-001 Lisboa, Portugal}



\begin{abstract}
Electron-positron pair cascades developed in the extreme electromagnetic fields of neutron star polar caps are considered a key source of magnetospheric plasma in these objects. We use a simplified model that maps the Quantum Electrodynamics processes governing the pair cascades to analytically and numerically model the development of the pair cascade, and show that large-amplitude oscillations of the electric field are inductively driven by the resulting plasma. A plasma instability arises in these oscillations, and particles accelerated in growing electric field perturbations can drive secondary pair bursts that damp the large-amplitude oscillations. An analytical model is proposed to describe this interplay between the pair production and kinetic collective plasma processes. All analytical results are shown to be in excellent agreement with particle-in-cell simulations.
\end{abstract}



\section{Introduction} \label{sec:intro}

Neutron star (NS) magnetospheres are thought to be filled with pair plasma generated in strong cascades in vacuum gaps. In these gaps, permeated by strong NS rotationally-induced electric fields, electrons and positrons are accelerated along magnetic field lines up to TeV energies, emitting gamma-ray curvature photons. These, in turn, can be reabsorbed in the extreme magnetic field of these objects ($B \sim 10^{12}$~G), producing pairs via Quantum Electrodynamics (QED) processes. The cascade process is interrupted when the current driven by the fresh pair plasma is able to screen the accelerating electric field. This requires a charge density comparable to the Goldreich-Julian~\citep{goldreich_julian_1969} (GJ) density $\rho_\textrm{GJ} \simeq \Omega B / 4 \pi c$, where $\Omega$ is the NS rotation frequency and $c$ is the speed of light. At the time the electric field is screened, the plasma carries a finite current, conducted by electrons and positrons previously accelerated in opposite directions. This plasma current reverses the initial electric field, which consecutively decelerates particles and reverses their momentum. This process is repeated, driving large-amplitude inductive oscillations~\citep{levinson_2005} which have been recently shown to couple to electromagnetic modes in the presence of a background magnetic field and identified as a possible source of radio emission in pulsars~\citep{philippov_2020}.

The production of strong QED cascades in vacuum gaps of rotating NS has long been proposed to be connected to coherent emission mechanisms from these objects~\citep{sturrock_1971, ruderman_sutherland_1975}. Many works have since attempted to analytically describe the vacuum gaps, e.g. focusing on observable signatures such as the shape of the pair production fronts~\citep{arons_1979, arons_1983}, or their ability to power emission depending on magnetospheric plasma conditions far from the gaps~\citep{beloborodov_2008}. Recent advances in numerical algorithms allowed numerical studies of NS vacuum gaps including the relevant QED effects from first principles in plasma kinetic simulations~\citep{timokhin_2010, timokhin_2013}, showing that the gaps open periodically, creating super-GJ density plasma bursts that screen the electric field driving the discharges, advecting then into the outer magnetosphere. The excitation of large-amplitude oscillations following the QED cascades has been studied using simplified heuristic models of pair production~\citep{levinson_2005, philippov_2020}, but a first-principles description of the model parameters has not established. Reduced models have also been used in global simulations of pulsar magnetospheres~\citep{chen_2014, philippov_2015}, and shown to be an efficient source of the magnetospheric plasma, even when limited to occur at low altitudes~\citep{chen_2020a}. Despite these efforts to model NS vacuum gaps (and others to model analogue gaps in black hole magnetospheres~\citep{levinson_2018, chen_2020b, kisaka_2020, crinquand_2020}) including \textit{ab initio} QED processes, the interplay between these and plasma kinetic processes remains poorly understood, and a theory that couples QED to the full temporal evolution of the plasma is missing.

In this work, we consider a simplified model for the emission and pair production processes and analytically describe the cascade process and the consequent self-consistent excitation and damping of current-driven plasma waves. All results obtained analytically are confirmed numerically using Particle-in-Cell (PIC) simulations performed with OSIRIS~\citep{fonseca_2002, fonseca_2008}. In section \ref{sec:model}, we formulate the pair production model and determine its mapping to a first-principles description. The growth rate of the cascade and the plasma distribution function at the end of the discharges in an idealized configuration of vacuum gaps are analytically determined in sections \ref{sec:grate} and \ref{sec:dist}, respectively. We find that the discharge sets the plasma in an oscillatory equilibrium, with the electric field oscillating in time with large amplitude. We describe these oscillations in section \ref{sec:oscillations}, and show that secondary pair creation bursts are triggered by particles accelerated in growing perturbations of the electric field. We show that these secondary bursts are responsible for damping the average electric field, and estimate the damping coefficient as a function of the model parameters in section \ref{sec:damping}. In section \ref{sec:corotating}, we discuss how the results derived in sections \ref{sec:grate} -- \ref{sec:damping} should be interpreted in the presence of a background current imposed by the global magnetosphere. Our conclusions are presented in section \ref{sec:conclusion}.

\section{Simplified pair production model} \label{sec:model}


We consider in this work pair discharges regulated by the QED processes of curvature radiation and pair production by absorption of gamma-ray photons in an intense magnetic field. The differential probability rates for these processes are well-known~\citep{erber_1966, ritus_1985} functions of the parent particles' energy, momentum components and local electromagnetic field components. These functions are complex, however they can be understood as critically depending on the quantum parameters $\chi_{\pm,\gamma}$ of the electrons, positrons (subscript $\pm$, respectively) and photons (subscript $\gamma$) involved in the processes. The quantum parameters can be defined, in general, as $\chi_{\pm, \gamma} = \sqrt{(p_\mu F^{\mu \nu})^2} / (B_Q m_e c^2)$, where $p^\mu$ is the four-momentum of the particle, $F^{\mu \nu}$ is the electromagnetic tensor, $m_e$ is the electron mass and $B_Q \simeq 4.4 \times 10^{13}$~G is the Schwinger critical field. When $\chi_{\pm,\gamma} \ll 1$, particles behave classically and QED effects may be disregarded. As $\chi_{\pm, \gamma}$ becomes sufficiently large, these effects become relevant. The exact value of $\chi_{\pm, \gamma}$ at which each process becomes relevant depends on its differential probability rate, e.g. for pair production it has an exponential cutoff for $\chi_\gamma \sim 0.1$~\citep{erber_1966}. We note that, for curvature radiation, it is possible to establish an equivalence with local synchrotron emission as derived from QED in the classical regime for highly relativistic particles~\citep{delgaudio_2020}. In this regime, the differential probability rate of curvature radiation matches that of QED synchrotron, with the critical energy $\varepsilon_\textrm{crit} \simeq 2 \chi_\gamma \varepsilon_\pm / 3 \chi_\pm^2$, where $\varepsilon_\pm$ is the energy of the emitting lepton.

For leptons, $\chi_\pm$ has a simple interpretation, as it measures the ratio between the field that particles experience in their rest frame and the Schwinger field, $B_Q \simeq 4.4 \times 10^{13}$~G. For photons, this interpretation is obviously invalid. However, we can interpret $\chi_\gamma$ for photons in a typical configuration in NS polar caps. In these settings, curvature photons are emitted along the local curved magnetic field. As they propagate in straight lines, an angle $\theta$ between their momentum vector and the local magnetic field builds up, and we can write $\chi_\gamma \simeq \theta (B / B_Q) (\varepsilon_\gamma / m_e c^2)$, where $\theta \ll 1$ and $\varepsilon_\gamma$ is the photon energy. The angle $\theta$ can be expressed as the ratio between the distance propagated by the photon and the magnetic field curvature radius, $\theta \simeq \ell_\gamma / \rho_C$. As $\ell_\gamma$ increases, so does $\chi_\gamma$ until the probability for pair production is non-negligible and the photon converts into a new electron-positron pair.

For a NS surface field $B\simeq 10^{12}$~G, and assuming that electrons/positrons move almost exactly along magnetic field lines, we expect $\chi_\pm \ll 1$, i.e. leptons emit photons classically~\citep{erber_1966}. In this process, an electron/positron with Lorentz factor $\gamma_\pm$ and quantum parameter $\chi_\pm$ produces a photon with energy $\varepsilon_\gamma$ and quantum parameter $\chi_\gamma$. The energies and quantum parameters of the parent and child particles can be related as $\varepsilon_\gamma \simeq \gamma_\pm m_e c^2 \chi_\gamma / \chi_\pm$. In general, the photon quantum parameter $\chi_\gamma$ has a distribution given by the differential probability rate for this process. However, in the classical emission regime, this distribution peaks sharply at $\chi_\gamma \simeq \chi_\pm^2$, and we can thus write $\varepsilon_\gamma \simeq \gamma_\pm m_e c^2 \chi_\pm$. This energy corresponds to the critical energy of a synchrotron type spectrum, with a trajectory radius $\sim \rho_C$.

Photons are typically emitted at $\chi_\gamma \ll 1$; however, as described above, during propagation they can probe regions of higher field intensity and/or curvature, and experience a $\chi_\gamma'$ that results in a non-negligible probability for pair production. If that occurs, the secondary pair particles are emitted with energy $\varepsilon_\pm'$ and quantum parameters $\chi_\pm'$, related to the parent particle properties as $\varepsilon_\pm' \simeq \varepsilon_\gamma' \chi_\pm' / \chi_\gamma'$. Since photons are emitted at $\chi_\gamma \ll 1$ and $\chi_\gamma$ build up with propagated distance, it is expected that pair production occurs at the lowest possible quantum parameter, which can be shown to be $\chi_\gamma' \simeq 0.1$~\citep{erber_1966}. In this range, the differential probability rate for pair production peaks at $\chi_\pm' \simeq \chi_\gamma' / 2$, and we can write $\gamma_\pm' m_e c^2 \simeq \varepsilon_\gamma' / 2$, i.e. the photon energy is equally split between the secondary particles. Given that photon energy does not change between the emission and pair production events, we can relate the Lorentz factor of secondary electrons/positrons with that of the primary lepton, $\gamma_\pm' \simeq (\chi_\pm/2) \gamma_\pm$.

In our model, any electron or positron that reaches a Lorentz factor $\gamma_\textrm{thr}$ emits a new pair with combined, equally split energy $\epsilon_\textrm{pair} = \gamma_\textrm{pair} m_e c^2$, where $\gamma_\textrm{pair} \equiv f \gamma_\textrm{thr}$. For simplicity, we assume that photons have zero mean free path, i.e. pair production is done in-place. The ratio between the primary and combined secondary particle energies is then $f = \gamma_\textrm{pair} / \gamma_\textrm{thr} = 2 \gamma_\pm' / \gamma_\pm \simeq \chi_\pm$. 

\section{Cascade growth rate} \label{sec:grate}

For realistic pulsar parameters, $f \simeq \chi_\pm \ll 1$, however we consider that a cascade can develop with arbitrary $f$. Assuming the cascade process develops in a region of uniform background field $E_0$, the plasma can be divided in two populations: 1) particles with a Lorentz factor $\gamma \in [\gamma_\textrm{thr}-\gamma_\textrm{pair},  \gamma_\textrm{thr}]$, and 2) secondary particles, with $\gamma \in [\gamma_\textrm{pair}/2, \gamma_\textrm{thr}-\gamma_\textrm{pair}[$. We define here two time scales relevant to describe this system. The first one is the time required for particles to be accelerated to $\gamma_\textrm{thr}$ starting from rest, and is defined as $t_a \equiv \gamma_\textrm{thr} m_e c / e E_0$, where $e$ is the elementary charge. The second one, $t_p \equiv f t_a$, is the time it takes for particles to be accelerated from $\gamma_\textrm{thr}-\gamma_\textrm{pair}$ to $\gamma_\textrm{thr}$, i.e. it is the period at which each particle in population 1 emits new pairs.

Particles in population 1 are never converted into population 2, however particles in population 2 created due to pair production are accelerated and get in the energy range of population 1 over a time $(1-3f/2)t_a = t_a - 3/2t_p$. We can thus write
\begin{equation}
n_1(t+(1-3f/2)t_a) = n_1(t) + (1-f) n_2(t) \ ,
\label{eq:cascade_delay}
\end{equation}
where $n_{1,2}$ are the number of particles in each population. The factor $(1-f)$ is only a small correction, and removes from the equality in equation~\eqref{eq:cascade_delay} the fraction of particles in population 2 at the time $t$ that decelerates and counter-propagates with the bulk distribution. This small fraction corresponds to electrons emitted by positrons (or vice-versa), and can be written as $(1-3f/2)/(1-f/2) \simeq 1 - f$ for $f \ll 1$.

The number of particles in population 2 increases due to pair production of both electrons and positrons in population 1, which happens every time interval $t_p = f t_a$, and decreases due to the conversion to population 1 mentioned above, so we can write
\begin{equation}
\frac{\mathrm{d}n_2}{\mathrm{d} t} = \frac{2 n_1}{f t_a} - \frac{n_2}{(1-3f/2)t_a} \ .
\label{eq:cascade_deriv}
\end{equation}

This system of equations can be solved by employing a Laplace Transform, or by simply assuming solutions of the type $\exp (\Gamma t)$. For $f \ll 1$, equations~\eqref{eq:cascade_delay} and \eqref{eq:cascade_deriv} reduce to
\begin{equation}
\label{eq:cascate_grate}
e^{\Gamma t_a} \simeq \frac{2}{f \Gamma t_a} \ .
\end{equation}
This equation has an exact solution given by the Lambert $W$ function~\citep{corless_1996}, $\Gamma t_a  = W (2/f)$, which can be approximated to $\Gamma t_a \sim \ln (2/f)$ for small $f$.

We confirmed this solution by performing one-dimensional PIC simulations where a uniform electron-positron plasma of density $n_0$ is subject to an initial uniform electric field $E_0 / (m_e c \omega_p / e) = 10^{7/2}$, where $\omega_p^2 = 4 \pi e^2 n_0/m_e$ is the plasma frequency associated with $n_0$. The simulation domain has a length $L/(c/\omega_p) = 10^{3/2}$ discretized in $N_L = 2000$ grid cells, and uses periodic boundary conditions. Only 1 particle/cell/species is initialized, however this number rapidly grows and at the end of the initial cascade we have over $10^4$ particles/cell/species. The time step is $\Delta t \omega_p = 10^{-7/2}$, a value chosen to well resolve the time scale $t_p = f t_a = f \gamma_\textrm{thr}/E_0$. In all simulations presented in this work, $\gamma_\textrm{thr} = 500$.

\begin{figure}[t]
\centering
\includegraphics[width=3.375in]{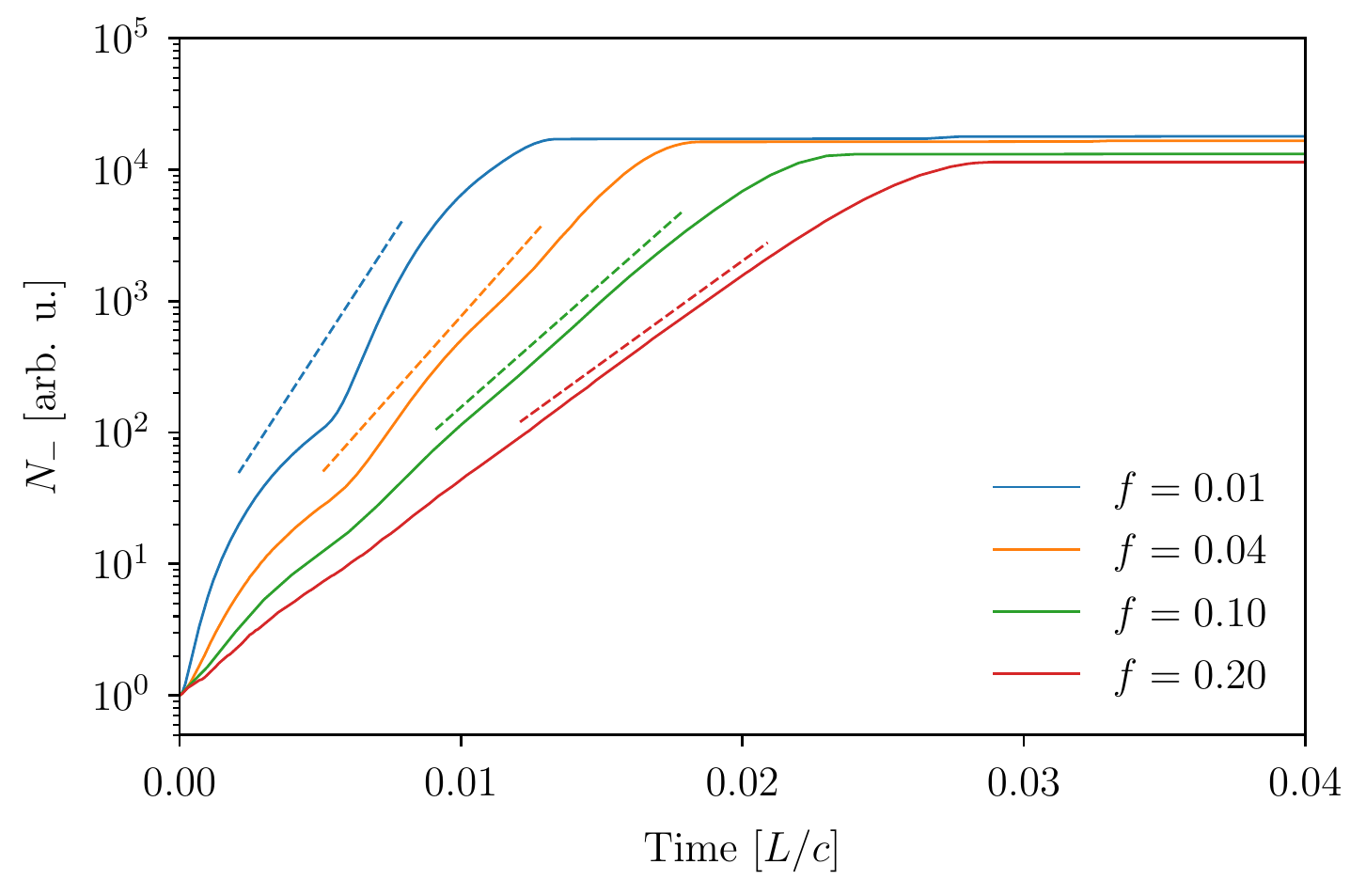}
\caption{\label{fig:ne_vs_time} Time evolution of number of electrons in cascades with different values of $f$ obtained in simulations (solid lines). The cascade growth rate is in excellent agreement with equation~\eqref{eq:cascate_grate} in all cases (dashed lines).}
\end{figure}

\begin{figure*}
\centering
\includegraphics[width=6.6in]{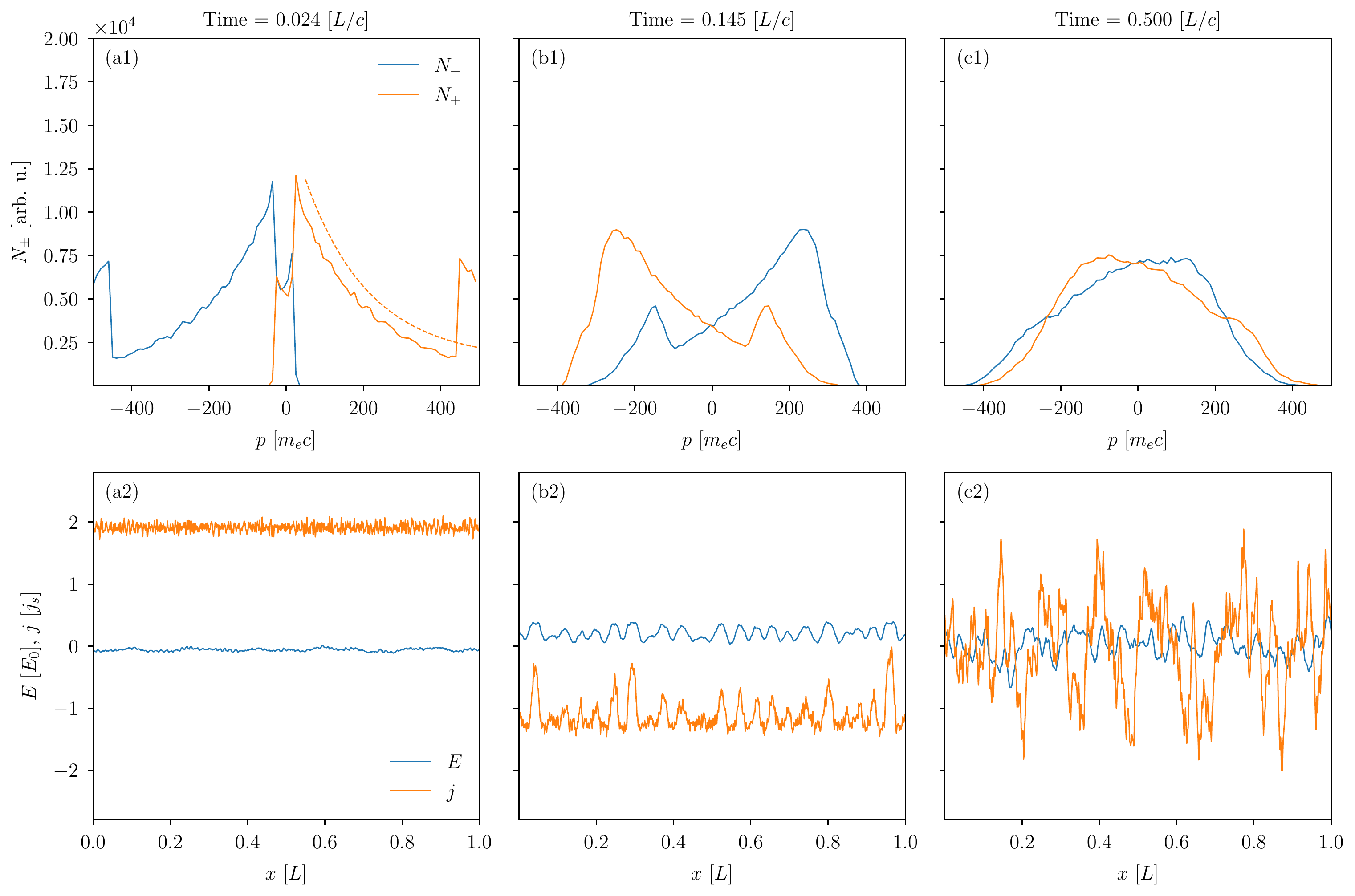}
\caption{\label{fig:phist_ej}Temporal evolution of particle distributions and average electric field damping following the initial cascade. The three columns correspond to times where (a) the initial cascade is fully developed, (b) electric field perturbations have grown and (c) the distribution function has reorganized in the nonlinear stage of the instability, corresponding to approximate times $\simeq 8$, $50$ and $175~\omega_0^{-1}$, respectively. The top row shows the momentum distribution of electrons and positrons, whereas the bottom row shows the electric field and current density (normalized to $j_s = e c n_s$) profiles for the same times. The dashed line in panel (1a) corresponds to the analytical distribution function in equation~\eqref{eq:dist_final}, and is drawn with an offset for clarity. Results obtained with $f=0.1$.}
\end{figure*}

The time evolution of the number of electrons $N_-$ in simulations with different values of $f$ is shown in Figure~\ref{fig:ne_vs_time}, confirming that the cascade grows faster for lower values of $f$. The cascade grows until the resulting plasma is dense enough to screen the electric field $E$. Amp\`{e}re's law then reads
\begin{equation}
\frac{\partial E}{\partial t} = -4 \pi j \simeq 4\pi e c (n_+ + n_-) \simeq 8\pi e c n_\pm \ ,
\label{eq:ampere_screen}
\end{equation}
where $j$ is the plasma current density and $n_\pm$ is the number density of positrons and electrons, respectively. Given that $n_\pm = n_0 \exp ( \Gamma t)$, we can solve equation~\eqref{eq:ampere_screen} to determine the time $t_s$ that it takes for the field to be screened, $t_s \simeq 1/\Gamma \ln (|E_0| \Gamma / (8 \pi e c n_0))$, a result consistent with previous works~\citep{levinson_2005}. This also allows us to compute the electron/positron number density at time $t_s$,
\begin{equation}
n_s \equiv n_\pm (t_s) = \frac{E_0^2}{8\pi m_e c^2} \frac{\ln (2/f)}{\gamma_\textrm{thr}} \ .
\label{eq:ns}
\end{equation}
In general, our simulations indicate that this expression sightly overestimates (by a factor of $1.5-2$) the plasma number density at the end of the cascade. This can be understood by noting that equation~\eqref{eq:ns} is derived assuming that the density grows exponentially in time in the cascade, which is not an exact description of $n_\pm(t)$ as it flattens towards the end of the process. Equation~\eqref{eq:ns} can also be read as an energy density balance that we discuss further in this manuscript.

\section{Plasma distribution function} \label{sec:dist}

During the cascade process, the plasma develops a broad energy distribution, with most of the particles in population 2. The distribution function of this population of positrons/electrons, $f_2$, can be determined by solving Vlasov equation,
\begin{equation}
\frac{\partial f_2}{\partial t} + \frac{eE_0}{m_e c} \frac{\partial f_2}{\partial \gamma} = 0 \ .
\label{eq:vlasov}
\end{equation}
We have assumed here that the cascade develops in a constant electric field $E_0$, and that all particles are relativistic, such that $p \simeq \gamma m_e c$. We emphasize also that this equation is only valid in the energy range of population 2, where only the electric field is responsible for energy transport, i.e. terms associated with pair creation and the conversion to population 1 are neglected.

Since the number of particles grows exponentially (with growth rate $\Gamma$), we can write the distribution function as $f_2(t,\gamma) = \bar{f}_2 \exp(\Gamma t) h(\gamma)$, where $\bar{f}_2$ is a normalization constant, and use equation~\eqref{eq:vlasov} to obtain $h(\gamma)$. The distribution function is then
\begin{equation}
f_2(t,\gamma) = \bar{f}_2 \exp(\Gamma t) \exp \left(- \frac{\Gamma m_e c}{eE_0} \gamma \right) .
\label{eq:dist}
\end{equation}
Using the definition of $t_a$ and the approximate solution for the growth rate $\Gamma t_a \simeq \ln (2/f)$, we can rewrite the result in equation~\eqref{eq:dist} as
\begin{equation}
f_2(t,\gamma) = \bar{f}_2 \exp(\Gamma t) \exp \left(- \ln(2/f) \gamma / \gamma_\textrm{thr} \right) .
\label{eq:dist_final}
\end{equation}
This expression holds only for energy ranges populated during the cascade, which can be seen by noting that the whole energy range is populated on a time scale $t_a$, whereas the cascade develops on a time scale of a few $t_c \sim 1/\Gamma$. The ratio of these two time scales is $t_a / t_c \simeq \ln (2/f)$, which depends only logarithmically on $f$ and is of order 10 for $f \lesssim 10^{-4}$, ensuring that the energy range $\gamma \in [1, \gamma_\textrm{thr}]$ is well populated before the field is screened. However, we note that even if this is not the case, the lower energy component of the distribution function is always occupied, and taking $f$ to decay exponentially with $\gamma$ for all energies according to equation~\eqref{eq:dist_final} is a robust assumption.

The normalization constant $\bar{f}_2$ can be calculated from the electron/positron number density at the time the field is screened, $n_s$. From equation~\eqref{eq:cascade_deriv}, we can write $n_1 / n_2 \simeq (f/2) \ln (2 /f)$ for $f \ll 1$, and thus $n_s = n_1(t_s) + n_2(t_s) \simeq n_2(t_s)$.  The full distribution function can then be expressed as $f(t, \gamma) \simeq f_2(t, \gamma) = \exp(\Gamma (t - t_s)) h'(\gamma)$, with the energy dependence written as
\begin{equation}
h'(\gamma) \simeq n_s \frac{\ln(2/f)}{\gamma_\textrm{thr}} \exp (-\ln(2/f)\gamma/\gamma_\textrm{thr}) .
\end{equation}

This distribution can be used to show that the average Lorentz factor of leptons is
\begin{equation}
\langle \gamma \rangle = \dfrac{\int \gamma h'(\gamma) \mathrm{d} \gamma}{\int h'(\gamma) \mathrm{d} \gamma} \simeq  \frac{\gamma_\textrm{thr}}{\ln(2/f)} \ .
\label{eq:gavg}
\end{equation}
As anticipated from the exponential decay of the distribution function with $\gamma$, this result shows that the average Lorentz factor is much lower than $\gamma_\textrm{thr}$. Combining equations~\eqref{eq:ns} and \eqref{eq:gavg}, we can also write $n_s \langle \gamma \rangle m_e c^2 = E_0^2 / (8 \pi)$, showing that the cascade process converts all energy density initially available into kinetic energy of electrons/positrons, with an average Lorentz factor $\langle \gamma \rangle$.

\section{Large-amplitude oscillations} \label{sec:oscillations}

In the previous section, we have shown that the cascade develops an electron-positron plasma with a broad energy distribution. In fact, electrons (positrons) develop a Lorentz factor distribution $\gamma \in [1, \gamma_\textrm{thr}]$ that extends below $\gamma_\textrm{pair}/2$ due to the emission by the counter-propagating positrons (electrons). At $t=t_s$, the plasma stops pair producing but continues driving a current, reversing the electric field. The plasma enters a regime where $E$ periodically oscillates in time, reversing the momentum of electrons and positrons and establishing large-amplitude plasma oscillations. The frequency of these oscillations can be derived from the time derivative of Amp\`{e}re's law,
\begin{equation}
\frac{\partial^2 E}{\partial t^2} = -4\pi \frac{\partial j}{\partial t} = - 8 \pi e \left ( v_+ \frac{\partial n_\pm}{\partial t} + n_\pm \frac{\partial v_+}{\partial t} \right) ,
\label{eq:ampere_full}
\end{equation}
where the current density has been written as $j = 2 e n_\pm v_+$, $v_+$ being the average velocity of positrons (electrons move with $v_- = -v_+$). Assuming that no pair production occurs, the first term on the right side of equation~\eqref{eq:ampere_full} can be dropped, and we can write
\begin{equation}
\frac{\partial^2 E}{\partial t^2} = - \frac{8 \pi e^2}{m_e} E \int \mathrm{d} p \ \frac{f_+(t, p)}{\gamma^3} = - \frac{8 \pi e^2 n_\pm}{m_e} \langle 1 / \gamma^3 \rangle E \ .
\label{eq:ampere_osc}
\end{equation}
Here, we have expressed the average positron velocity as $v_+ = \int \mathrm{d} p \ v f_+(t, p) / \int \mathrm{d} p \ f_+(t, p)$, where $f_+$ is the positron distribution function, normalized as $\int \mathrm{d} p \ f_+ = n_+$. We have also used Vlasov equation to write $\partial f_+ / \partial t = - e E \partial f_+/\partial p$, and then performed an integration by parts. Equation~\eqref{eq:ampere_osc} shows that the frequency of large-amplitude plasma oscillations is $\omega_0 = \sqrt{8 \pi e^2 n_s / m_e \langle 1/\gamma^3\rangle}$. The distribution function in equation~\eqref{eq:dist_final} can be used to compute $\langle 1 / \gamma^3 \rangle \simeq 1 / (2 \gamma_\textrm{thr})$. In this new equilibrium, both electron and positron momentum distributions slide exactly between $\pm \gamma_\textrm{thr} m_e c$ without producing new pairs in a non current neutral, oscillatory counter-streaming configuration. We find, however, that this equilibrium is unstable, and perturbations grow on the initially uniform electric field, accelerating particles in the high energy tail of the distribution to Lorentz factors beyond $\gamma_\textrm{thr}$, and thus driving secondary pair bursts. Hence, the high energy tail of the distribution creates new pairs that populate its bulk component, an interplay previously identified in QED cascades developed in ultra-intense, counterpropagating laser pulses~\citep{grismayer_2016, grismayer_2017}. Electric field perturbations grow until they become comparable to the electric field oscillation amplitude $E_0' \simeq E_0 / 2$, and the average electric field is damped. Figure~\ref{fig:phist_ej} illustrates the particle momentum distributions and the electric field and current density profiles at three distinct times during this process.

The instability that disrupts the equilibrium set after the cascade develops has a typical growth rate $\Gamma_i / \omega_0 \sim  0.1$, and has an infinite number of growing modes, the fastest being $k c / \omega_0 \sim 1 - 10$, as measured from simulations with different parameters. We find that $\Gamma_i$ is larger for lower $f$, and, for $f \lesssim 0.01$, the instability develops on the first oscillation cycle of the electric field, immediately damping the field. In this work, we focus our analysis in a simulation with $f=0.1$, where the oscillating and damping phases of the electric field are easily distinguishable. We find that this instability develops when multiple plasma species support a large-amplitude, oscillating background electric field, in both classical and relativistic regimes with either cold or warm distributions. Due to its generality and involved analytical description, a complete theoretical and numerical analysis of this instability will be presented elsewhere. Here, we focus on the implications of this instability in the electrodynamics of pair discharges, namely its role in triggering new pair production events.

\section{Average electric field damping} \label{sec:damping}

\begin{figure}[t]
\centering
\includegraphics[width=3.375in]{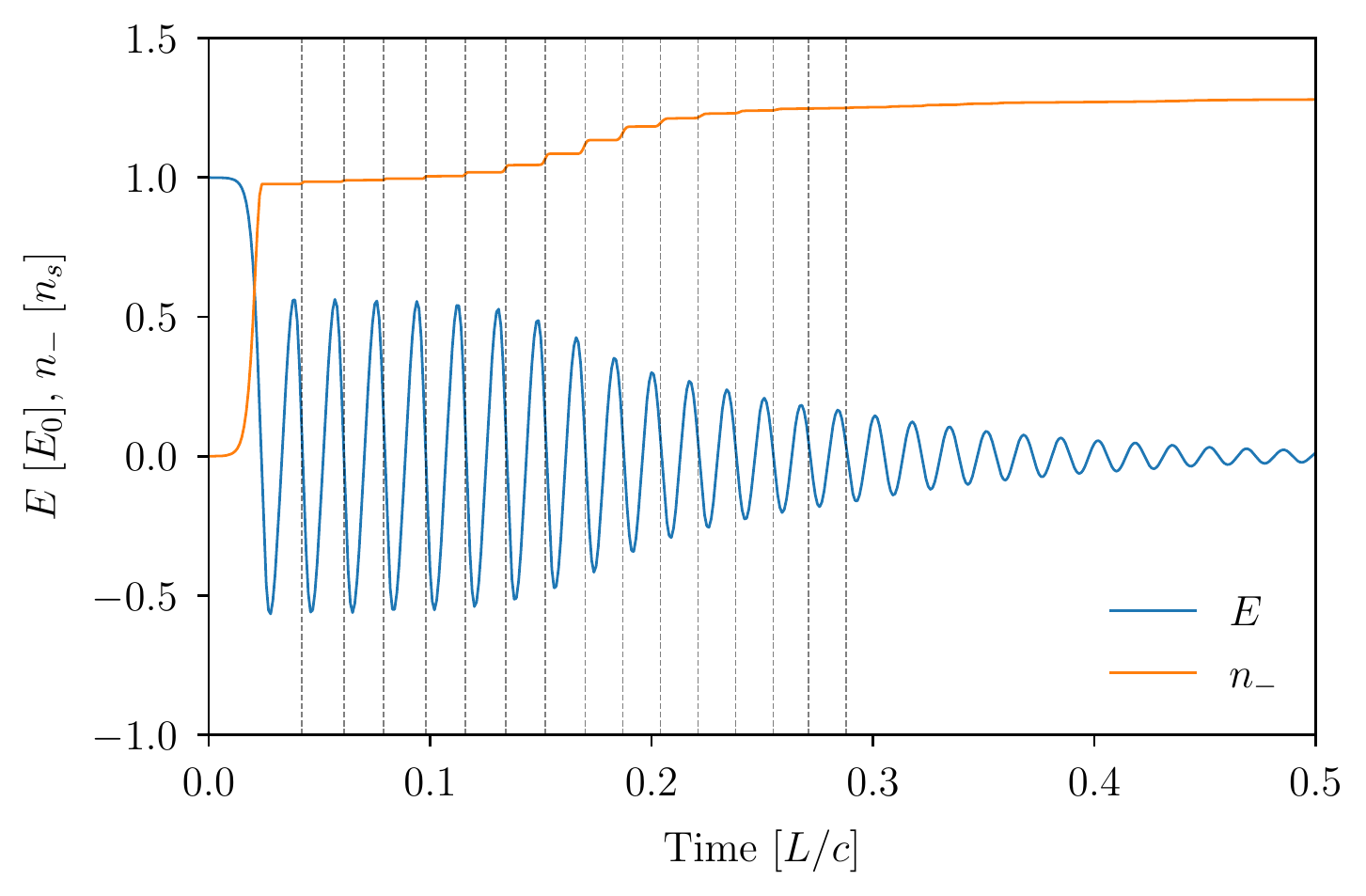}
\caption{\label{fig:ne_edamp} Damping of average electric field and growth of electron number density due to secondary pair bursts in simulation with $f=0.1$. The vertical dashed lines represent times where the average electric field is zero and secondary bursts occur.}
\end{figure}

Electrons and positrons accelerated in the perturbations grown on top of the uniform, oscillating electric field are able to produce secondary pair bursts, effectively damping it. We note that these secondary bursts occur in a well defined phase of the oscillation, as illustrated by the vertical dashed lines in Figure~\ref{fig:ne_edamp}. In particular, we observe new discharges when the average electric field is null, i.e. when the work done by the electric field on the particles in the previous (perturbed) half cycle is maximum. The time evolution of the electric field can be determined by solving equation~\eqref{eq:ampere_full}, considering now that the density $n_\pm$ changes only periodically and very suddenly at times $t_n$ where $v_+ \simeq c$ is constant. In these conditions, the second term on the right side of equation~\eqref{eq:ampere_full} can be neglected around $t=t_n$, and we obtain 
\begin{equation}
\frac{\partial^2 E}{\partial t^2} = - 8 \pi e c \frac{\partial n_\pm}{\partial t} = - \frac{m_e c}{e \langle 1 / \gamma^3 \rangle} \frac{\partial \omega^2(t)}{\partial t} \ ,
\label{eq:ampere_damp}
\end{equation}
where we have written $\omega^2(t) = (8 \pi e^2 / m_e ) \langle 1 / \gamma^3 \rangle n_\pm(t)$ and assumed that the average $\langle 1 / \gamma^3 \rangle$ is approximately constant during the sudden changes in density. A solution for equation~\eqref{eq:ampere_damp} can be obtained by assuming that secondary bursts develop on a time scale much shorter than $1/\omega_0$ and therefore can be modelled as step functions in the density,
\begin{equation}
\omega^2 (t) = \omega_0^2 + \sum_n \Delta^2_n H(t-t_n)\ .
\label{eq:omega2}
\end{equation}
where $\Delta_n^2$ is the amplitude of the $n$th density jump and $H(t)$ is the Heaviside function. We assume that the solution in each interval $t \in [t_{n-1}, t_n]$ is, in general,
\begin{equation}
\label{eq:e_gensol}
E(t) = E_n \cos(\omega_n t + \phi_n) \equiv E_n \cos \Phi_n \ .
\end{equation}
Integrating equation~\eqref{eq:ampere_damp} between $t_n^-$ and $t_n^+$, where $t_n^\pm = t_n \pm \delta t$ and $\delta t \ll 2\pi/\omega_0$, we obtain
\begin{equation}
\left. \frac{\partial E}{\partial t} \right|_{t_n^-}^{t_n^+} = - \frac{m_e c}{e \langle 1/\gamma^3 \rangle} \Delta_n^2 \ .
\label{eq:dedt_jump}
\end{equation}
We can perform a similar integration over Amp\`{e}re's law to obtain
\begin{equation}
E \bigg|_{t_n^-}^{t_n^+} = - \frac{m_e c}{e \langle 1/\gamma^3 \rangle} \left( \omega_0^2 + \frac{\Delta_n^2}{2} \right) \left( t_n^+ - t_n^- \right) \ .
\label{eq:e_jump}
\end{equation}
Taking the limit $\delta t \to 0$ in equation~\eqref{eq:dedt_jump} and \eqref{eq:e_jump}, we see that $E$ is continuous but its derivative has a finite jump at $t_n$ proportional to $\Delta_n^2$. Applying the solution in equation~\eqref{eq:e_gensol} to the jump conditions derived here, we obtain
\begin{subequations}
\begin{equation}
E_{n+1} \cos \Phi_{n+1} - E_n \cos \Phi_n = 0 \ ,
\end{equation}
\begin{equation}
\omega_{n+1} E_{n+1} \sin \Phi_{n+1} - \omega_n E_n \sin \Phi_n = \frac{m_e c}{e \langle 1/\gamma^3 \rangle} \Delta_n^2 \ .
\label{eq:dedt_jump_full}
\end{equation}
\end{subequations}
Noting that in the late stages of the cascade development to which the model here applies, the sudden increases in density occur when the electric field is zero, we can write
\begin{equation}
\cos \Phi_{n+1} = \cos \Phi_n = 0 \ ,
\end{equation}
which implies $\sin \Phi_{n+1} = \sin \Phi_n = \pm 1$ . Using this result, we can rewrite equation~\eqref{eq:dedt_jump_full} as
\begin{equation}
\frac{E_{n+1}}{E_n} = \frac{1}{\sqrt{1+\Delta_n^2/\omega_n^2}} + \frac{m_e c}{e \langle 1/\gamma^3 \rangle \omega_n} \frac{\Delta_n^2}{\sqrt{1+\Delta_n^2/\omega_n^2}} \frac{1}{E_n} \ ,
\label{eq:e_amp_frac}
\end{equation}
where we have also used the relationship $\omega_{n+1} = \omega_n \sqrt{1+\Delta_n^2/\omega_n^2}$ (from equation~\eqref{eq:omega2}). Writing $\Delta_n^2 = \epsilon_n \omega_n^2$, equation~\eqref{eq:e_amp_frac} can be rewritten as
\begin{equation}
E_{n+1} - E_n = \left[ \frac{1}{\sqrt{1+\epsilon_n}} - 1 \right] E_n + \frac{m_e c}{e \langle 1/\gamma^3 \rangle} \frac{ \epsilon_n \omega_n} {\sqrt{1+\epsilon_n}} \ .
\end{equation}
We now look for a continuous function, $E(t)$, that matches the series at every time $t_n$. To simplify the analysis, we consider that changes in the oscillation frequency are negligible, $\omega_n = \omega_0$, such that $t_n = 2\pi n /\omega_0$. In this case, we can write
\begin{equation}
\frac{\mathrm{d} E}{\mathrm{d} t} = \frac{\omega_0}{2\pi} \left[ \frac{1}{\sqrt{1+\epsilon(t)}} - 1 \right] E(t) + 
\frac{\omega_0}{2\pi} \frac{m_e c \omega_0 }{e \langle 1/\gamma^3 \rangle} \frac{ \epsilon (t) } {\sqrt{1+\epsilon(t)}} \ .
\label{eq:cont_dedt}
\end{equation}
Let us consider the case in which $\epsilon(t) \ll 1$. In this case, we can Taylor expand both denominators on the right-hand side of equation~\eqref{eq:cont_dedt} to get (to first order in $\epsilon$)
\begin{equation}
\frac{\mathrm{d} E}{\mathrm{d} t} = -\frac{\omega_0 \epsilon(t)}{4\pi}  E(t) + 
\frac{\omega_0 \epsilon(t)}{2\pi} \frac{m_e c \omega_0 }{e \langle 1/\gamma^3 \rangle} \ .
\label{eq:dedt_damp}
\end{equation}
Assuming, for simplicity, that the fraction of particles that convert into new pairs in each electric field cycle is constant in time, $\epsilon (t) = \epsilon_0$, the solution of equation~\eqref{eq:dedt_damp} is
\begin{equation}
E(t) = E_0' \exp \left(-\frac{\epsilon_0 \omega_0 t}{4 \pi} \right) + \frac{2 m_e c \omega_0 }{e \langle 1/\gamma^3 \rangle} \left[ 1 - \exp \left(-\frac{\epsilon_0 \omega_0 t}{4\pi} \right) \right] \ ,
\label{eq:e_damp}
\end{equation}
i.e. the electric field is exponentially suppressed, as observed in several previous works modelling pair cascades using both reduced~\citep{levinson_2005, philippov_2020} and first-principles~\citep{timokhin_2010, levinson_2018, chen_2020b, kisaka_2020} models of the QED processes. We note that assuming that $\epsilon$ is constant in time is unquestionably an oversimplification. In fact, $\epsilon(t)$ is controlled by the fraction of particles that can be accelerated to energies above $\gamma_\textrm{thr}$, given an electric field of amplitude $\langle E \rangle + \delta E$, where $\delta E$ is its exponentially growing perturbative component. Even if this can be calculated for times shortly after the field is initially screened by taking the distribution function in equation~\eqref{eq:dist_final} and computing the fraction of particles with $\gamma \in [\gamma_\textrm{thr} - \delta \gamma, \gamma_\textrm{thr}]$, where $\delta \gamma \sim e \delta E / m_e c$, this estimate would rapidly be invalid, as it is a complex function of the exponentially growing $\delta E$ and decaying $\langle E (t) \rangle$. In general, we observe that $\epsilon(t)$ initially increases with time as perturbations in the electric field grow, and then decreases due to the damping of the average electric field, as shown in Figure~\ref{fig:ne_vs_time}. We have verified that the solution in equation~\eqref{eq:e_damp} is robust to other temporal profiles of $\epsilon(t)$ by numerically solving equation~\eqref{eq:dedt_damp}. The decay seems to be roughly insensitive to changes in $\epsilon$; however, the residual amplitude of the field is controlled by the time scale $\tau$ on which $\epsilon$ vanishes, for instance if $\epsilon(t) = \epsilon_0$ for $t < \tau$ and zero otherwise, then the final amplitude of the average electric field is $E(\tau) \simeq E_0' \exp ( - \epsilon_0 \omega_0 \tau / 4 \pi)$, which, in general, largely exceeds the residual value $m_e c \omega_0/e \langle 1/\gamma^3 \rangle$ given in equation~\eqref{eq:e_damp}.

\section{Co-rotating frame} \label{sec:corotating}
In the previous sections, we have assumed that the plasma is unmagnetized. In general, pulsar polar caps are permeated by a very strong magnetic field that forbids electrons and positrons to cross field lines, and a one dimensional description of the plasma (along the magnetic field lines) in this region is enough, provided that photons decay within a small distance. However, the global magnetosphere imposes a local current in pulsar polar caps that supports a twist in the open field lines with footpoints within the polar cap~\citep{arons_1979}. In this section, we discuss how are the results presented before modified in the presence of a background current imposed by the magnetosphere. To illustrate possible differences, we have performed a one-dimensional PIC simulation in a frame co-rotating with the NS that correctly accounts for the current imposed by the global magnetosphere, $j_\textrm{m}$. This simulation is similar to those presented e.g. in \citet{timokhin_2010}. We outline here the simulation setup and describe its parameters for completeness.

In the frame co-rotating with the NS, Gauss' law reads
\begin{equation}
\frac{\partial E}{\partial x} = 4 \pi (\rho - \rho_\textrm{GJ}) \ ,
\label{eq:gauss_corotating}
\end{equation}
whereas Faraday's law is written as
\begin{equation}
\frac{\partial E}{\partial t} = - 4 \pi (j - j_\textrm{m}) \ .
\label{eq:ampere_corotating}
\end{equation}
The simulation presented in this section adopts the form of Faraday's law in equation~\eqref{eq:ampere_corotating}, with $j_\textrm{m} = - 1.5 \rho_\textrm{GJ} c$. The exact value of $j_\textrm{m} / \rho_\textrm{GJ} c$ controls the efficiency of particle acceleration in the polar cap and of pair production. A discussion of the dependence of gap properties on the exact value of $j_\textrm{m}$ can be found e.g. in \citet{beloborodov_2008, timokhin_2010}. We also consider here that $\rho_\textrm{GJ} < 0$. The simulation domain has a length $L_x / (c / \omega_{p,\textrm{GJ}}) = 1000$, where $\omega_{p, \textrm{GJ}}^2 = 4 \pi e^2 n_\textrm{GJ} / m_e$ and $n_\textrm{GJ} = |\rho_\textrm{GJ}| / e$. The number of cells is such that the grid size is $\Delta x / (c / \omega_{p,\textrm{GJ}}) = 0.1$, and the time step is $\Delta t \omega_{p, \textrm{GJ}} = 10^{-3}$. Open boundary conditions are used for both fields and particles. Electrons and positrons are initially distributed uniformly in space, with charge densities $\rho_- = - 2 |\rho_\textrm{GJ}|$ and $\rho_+ = |\rho_\textrm{GJ}|$, respectively. This ensures that equation~\eqref{eq:gauss_corotating} is satisfied by the initial electric field, $E = 0$. Both species are initialized with a cold velocity distribution such that $j = j_\textrm{m}$, ensuring also that equation~\eqref{eq:ampere_corotating} is satisfied. Pair production is governed by the energy-based model described in sections \ref{sec:model} and \ref{sec:grate}, with $\gamma_\textrm{thr} = 5000$ and $f = 0.01$.

When the simulation starts, the electric field remains zero everywhere except close to the simulation boundaries. As shall become clear, the plasma dynamics close to the boundaries is nearly symmetric, so we will focus on the left boundary. We will refer to this boundary as the NS surface, and to the positive $x$ direction as pointing towards the outer magnetosphere. As positrons leave through this boundary, and electrons move away from the surface, the vacuum gap electric field develops. Positrons are accelerated in the gap towards the NS surface until they reach a Lorentz factor $\gamma_\textrm{thr}$. As pairs are created, secondary positrons are absorbed at the surface, and a beam of electrons is accelerated towards the magnetosphere. The beam of electrons is accelerated in the unscreened electric field until it starts emitting pairs. The positrons emitted at this point are then accelerated backwards towards the star, but as they do, the plasma trail that follows the electron beam reverses its current density sign, and so does the electric field. At this point, this trail plasma is in the configuration obtained following the initial cascade in the model presented in section \ref{sec:oscillations}. Even if the trail plasma has broad energy distribution, in general it does not extend to $\gamma_\textrm{thr} m_e c^2$. However, as the electric field is continuously reversed following the electron beam, parts of this beam can be accelerated to energies beyond $\gamma_\textrm{thr} m_e c^2$, producing secondary bursts of pairs. The role of these consecutive bursts on damping the electric field following the electron beam is qualitatively similar to the model presented in \ref{sec:damping}. As the beam and trailing plasma move away from the surface, the gap develops again and the process is repeated.

\begin{figure*}
\centering
\includegraphics[width=5.76in]{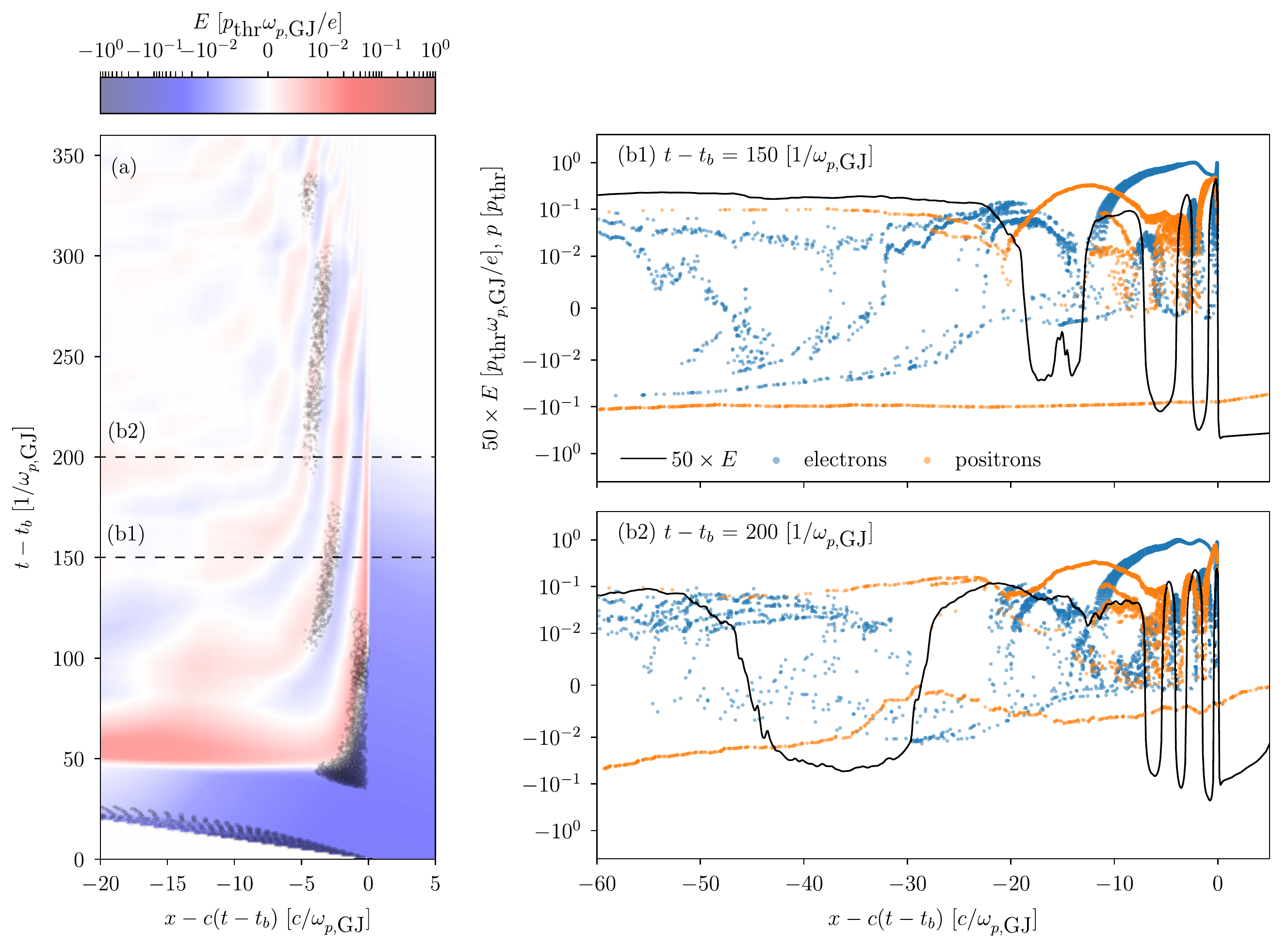}
\caption{\label{fig:corotating} Spatiotemporal plasma dynamics following a burst of pair production. Panel (a) shows the electric field as a function of space and time in colors. Black circles in this panel represent pair production events, and the size of the circle is proportional to the number of events in a given temporal and spatial coordinate. Panels (b1) and (b2) show the electric field profile in black and the electron and positron phase spaces in blue and orange respectively. Particle momenta have been normalized to $p_\textrm{thr} = \gamma_\textrm{thr} m_e c$.}
\end{figure*}

Figure~\ref{fig:corotating} illustrates the spatiotemporal plasma dynamics following a single burst of pairs triggered at time $t = t_b$. Panel (a) shows the electric field as a function of time and space. The spatial coordinate moves with the velocity of the burst, $v_b = c$, such that its front appears to be stationary. Black circles in panel (a) represent pair production events. Pairs emitted for $(t - t_b) \omega_{p, \textrm{GJ}} \lesssim 25$ are produced by the initial positron beam accelerated towards the surface. All other pairs are emitted by electrons in the secondary beam. As the leading front of this beam is accelerated past $\gamma_\textrm{thr} m_e c^2$, a large number of pairs is created, which happens for $(t - t_b) \omega_{p, \textrm{GJ}}$ between $\sim 50$ and $\sim 100$. As the field is reversed, part of the electron beam following the first $E$ peak can be further accelerated, and produces a second burst of pairs, delayed in time and space from the first one ($(t - t_b) \omega_{p, \textrm{GJ}}$ between $\sim 100$ and $\sim 170$). This occurs repeatedly for further delayed positions in the electron beam, and the electric field is damped. Panel (a) also shows that the oscillations driven in the electric field by the repeated reversals in plasma current travel superluminally, as observed in other works~\citep{timokhin_2010, philippov_2020}. Panels (b1) and (b2) show the electric field profile and the electron and positron phase spaces at different times. The exponential decay is well illustrated in these panels, as well as the peaked energy distribution of the electron beam in consecutive positions that gives rise to the regular secondary pair bursts. Panels (b1) and (b2) also show the oscillatory character of the electric field far downstream from the burst location, as described in section \ref{sec:oscillations}.

\section{Conclusion} \label{sec:conclusion}

The analysis presented in this work provides a complete picture of the development of pair cascades in NS polar caps, including, for the first time, a description of the interplay between QED and plasma kinetic processes. We present an analytical model of the initial cascade and associated screening of the electric field, and we show that an oscillatory equilibrium is set, with the electric field oscillating with high amplitude at the relativistic plasma frequency. This equilibrium is shown to be unstable, and particles accelerated in the electric field perturbations produce secondary pair bursts, redistributing the energy in the high energy tails of the distribution function to its bulk. We have analytically demonstrated that the repeated creation of new pairs damps the electric field, and that the final amplitude of the electrostatic oscillations is determined by how quickly the instability disrupts the oscillatory equilibrium. Our description is based on simplified approximations to realistic models, in particular regarding the QED processes governing pair cascades. However, we have shown that the model parameter $f$, controlling the separation between the energy of primary particles and losses to curvature radiation, is a proxy for the quantum parameter of primary particles, $f \simeq \chi_\pm$. Under realistic conditions ($B = 10^{12}$~G and a NS rotation period of $T = 1$~s), we can estimate $\chi_\pm \simeq 10^{-6}$, assuming that photons produced at the NS surface decay within a distance of $10$~m, i.e. a fraction of the extent of the polar cap vacuum gap. This estimate can also be used to compute $\gamma_\textrm{thr} \simeq 7.3 \times 10^{7}$, which is a free parameter in this model. In numerical simulations, $f \simeq 10^{-6}$ is not achievable. However, we have verified that the analysis presented in this work holds for values as low as $f = 10^{-3}$. The model also does not assume any spread in $\gamma_\textrm{thr}$ and pair production is done in-place, which may play a role in smoothing the plasma distribution function developed in the cascade. We have also considered the initial plasma seed and vacuum gap electric field to be uniform in space. In reality, the electric field is expected to decay with altitude within a distance comparable to the polar cap radius. Nevertheless, we do not expect either of these approximations to significantly change the main conclusions drawn in this work in more realistic models. Furthermore, we have shown that the results obtained in an idealized configuration hold qualitatively in simulations performed in a co-rotating frame. Large-amplitude oscillations of the electric field are also observed in these simulations. Secondary pair production bursts are driven in this case by an electron beam accelerated in the large-amplitude electric field oscillations. The consecutive bursts exponentially damp the field from the head to the tail of the beam. Inductive plasma oscillations are also observed to develop downstream from the burst. We expect these oscillations to be potential sites for linear acceleration emission. In particular, we observe particle trajectories to be similar to the oscillatory orbits described in~\citet{melrose_2009a, reville_2010} and shown to complement curvature radiation in the production of high energy photons when the acceleration length is smaller than the formation length of curvature photons.

The results presented in this work are particularly relevant for other studies attempting to model the microphysics of vacuum gaps using multi-dimensional simulations with heuristic descriptions of the pair production processes, as the mapping between the reduced and \textit{ab initio} descriptions of the relevant QED processes provided here supports its applicability. Moreover, we expect that the analysis presented here holds qualitatively in multi-dimensional scenarios, given that the only assumption on the electromagnetic field configuration in our work is that there is an initial background electric field. The electric field considered here is that responsible for accelerating the particles until pair production, and so in NS polar caps, this would be the component parallel to the magnetic field. In fact, the onset and damping of large-amplitude inductive oscillations are observable in results presented in recent works with two-dimensional simulations~\citep{philippov_2020}. Furthermore, we conjecture that the interplay between the QED and plasma kinetic processes outlined here may also provide a good baseline understanding of other gaps in NS and black hole magnetospheres (governed by these or other QED processes) where the abstraction from the details of the QED mechanisms may be applicable.

\acknowledgments

This work was supported by the European Research Council (ERC-2015-AdG Grant 695088) and FCT (Portugal) (grant PD/BD/114307/2016) in the framework of the Advanced Program in Plasma Science and Engineering (APPLAuSE, FCT grant PD/00505/2012). We acknowledge PRACE for granting access to MareNostrum, Barcelona Supercomputing Center (Spain), where the simulations presented in this work were performed.


\bibliography{sample63}{}
\bibliographystyle{aasjournal}



\end{document}